\documentclass[%
 amsmath,amssymb,
 aps,
prd,showpacs,onecolumn,superscriptaddress
]{revtex4-2} 

\usepackage{graphicx} 
\usepackage{subfigure} 
\usepackage{dcolumn} 

\usepackage{amsmath} 
\usepackage{amssymb} 
\usepackage{bm} 
\usepackage{color}

\usepackage{url} 
\usepackage[normalem]{ulem} 
\usepackage[dvipsnames]{xcolor} 
\usepackage{hyperref} 
\usepackage{bm} 

\usepackage{amsthm,amsmath,amssymb} 
\usepackage{mathrsfs} 
\usepackage{url} 
\def\be{\begin{equation}}
\def\ee{\end{equation}}
\def\bea{\begin{eqnarray}}
\def\eea{\end{eqnarray}}
\newcommand{\bes}{\begin{subequations}}
\newcommand{\ees}{\end{subequations}}
\def\comment#1{}

\allowdisplaybreaks

\begin{document}

\title{Detecting extreme-mass-ratio inspirals for space-borne detectors with deep learning}

\author{Qianyun Yun}

\affiliation{Shanghai Astronomical Observatory,  Chinese Academy of Sciences,  Shanghai,  China,  200030}
\affiliation{School of Physics and Astronomy, Shanghai Jiao Tong University 800 Dongchuan RD.,Minhang District, Shanghai, 200240, China}

\author{Wen-Biao Han}
\email{Corresponding author: wbhan@shao.ac.cn}

\affiliation{Shanghai Astronomical Observatory,  Chinese Academy of Sciences,  Shanghai,  China,  200030}
\affiliation{Hangzhou Institute for Advanced Study, University of Chinese Academy of Sciences, Hangzhou 310124, China}
\affiliation{School of Astronomy and Space Science,  University of Chinese Academy of Sciences,  Beijing,  China,  100049}
\affiliation{Taiji Laboratory for Gravitational Wave Universe (Beijing/Hangzhou), University of Chinese Academy of Sciences, Beijing 100049, China}

\author{Yi-Yang Guo}
\affiliation{Lanzhou Center for Theoretical Physics, Key Laboratory of Theoretical Physics of Gansu Province,\\
and Key Laboratory of Quantum Theory and Applications of MoE,\\
Lanzhou University, Lanzhou, Gansu 730000, China
}
\affiliation{Institute of Theoretical Physics \& Research Center of Gravitation, Lanzhou University, Lanzhou 730000, China}

\author{He Wang}
\affiliation{International Centre for Theoretical Physics Asia-Pacific (ICTP-AP), University of Chinese Academy of Sciences (UCAS), Beijing, China
}
\affiliation{Taiji Laboratory for Gravitational Wave Universe (Beijing/Hangzhou), University of Chinese Academy of Sciences, Beijing 100049, China}

\author{Minghui Du}
\affiliation{ Center for Gravitational Wave Experiment, National Microgravity Laboratory, Institute of Mechanics, Chinese Academy of Sciences, Beĳing 100190, China}

\bibliographystyle{plain}

\begin{abstract}
One of the primary objectives for space-borne gravitational wave detectors is the detection of extreme-mass-ratio inspirals (EMRIs). This undertaking poses a substantial challenge because of the complex and long EMRI signals, further complicated by their inherently faint signal. In this research, we introduce a 2-layer Convolutional Neural Network (CNN) approach to detect EMRI signals for space-borne detectors. Our method employs the Q-transform for data preprocessing, effectively preserving EMRI signal characteristics while minimizing data size. By harnessing the robust capabilities of CNNs, we can reliably distinguish EMRI signals from noise, particularly when the signal-to-noise~(SNR) ratio reaches 50, a benchmark considered a ``golden''  EMRI. At the meantime, we incorporate time-delay interferometry (TDI) to ensure practical utility.  We assess our model's performance using a 0.5-year dataset, achieving a true positive rate~(TPR) of 94.2\% at a 1\% false positive rate~(FPR) across various signal-to-noise ratio form 50-100, with 91\% TPR and 1\% FPR at an SNR of 50. This study underscores the promise of incorporating deep learning methods to advance EMRI data analysis, potentially leading to rapid EMRI signal detection.\\

\end{abstract}
\maketitle

\section{Introduction}
Since the initial detection of GW150914~\cite{LIGOScientific:2016dsl, LIGOScientific:2014pky, LIGOScientific:2016aoc, LIGOScientific:2018urg}, ground-based gravitational wave detectors such as LIGO~\cite{Harry:2010zz} and VIRGO~\cite{VIRGO:2014yos}  have made remarkable progress. These detectors have subsequently detected nearly one hundred analogous events of the merger of two black holes with stellar mass~\cite{LIGOScientific:2018mvr, LIGOScientific:2020ibl, LIGOScientific:2021usb, LIGOScientific:2021djp}. These observations offer a fresh outlook on the origin and development of these celestial bodies~\cite{LIGOScientific:2020kqk, KAGRA:2021duu, Belczynski:2017gds}, enabling researchers to test Einstein's general theory of relativity~\cite{LIGOScientific:2019fpa, LIGOScientific:2021djp, LIGOScientific:2021sio} and explore the universe through an independent approach~\cite{LIGOScientific:2021aug}.

Ground-based gravitational wave detectors such as LIGO, Virgo, and Kagra~\cite{Aso:2013eba, KAGRA:2013rdx} have a primary mission of focusing on gravitational waves within a frequency range of 1 Hz to a few thousand Hz, as determined by their sensitivity curve~\cite{LIGOScientific:2021djp, Kaiser:2020tlg}. In contrast, space-based gravitational wave detection projects, represented by missions like LISA~\cite{LISA:2017pwj}, ASTROD~\cite{Ni:2012eh}, DECIGO/BBO~\cite{Kawamura:2011zz}, ALIA~\cite{Crowder:2005nr}, AGIS-LEO~\cite{Hogan:2011tsw}, Taiji~\cite{Hu:2017mde, Ruan:2018tsw}, and Tianqin~\cite{TianQin:2015yph}, are specifically engineered to detect low-frequency gravitational waves, covering the spectrum from 0.1 mHz to 1 Hz~\cite{LISA:2017pwj, Hu:2017mde, Kaiser:2020tlg}. These missions are primarily focused on exploring a wide range of phenomena related to black holes, compact stars, and the origins of the universe. They pay special attention to topics like  massive black hole binaries (SMBH)~\cite{Klein:2015hvg}, extreme-mass-ratio inspirals ~\cite{Babak:2017tow, Han:2018hby, Xin:2018urr, Yang:2019xro}, binary white dwarfs (BWD)~\cite{Lamberts:2019nyk} the potential presence of a stochastic gravitational wave background (SGWB)~\cite{Caprini:2015zlo, Bartolo:2016ami} and so on.

As early as 2008, China initiated the Taiji project, which consists of three spacecrafts. These spacecrafts follow heliocentric orbits and come together to form a massive equilateral triangle with sides spanning approximately three million kilometers~\cite{Gong:2011zzd}. The Taiji space-based detectors are positioned at a distance of about 1 astronomical unit (1AU) from the Sun. The center of mass of this constellation either lags or leads the Earth by approximately 18 to 20 degrees.  The eccentricity of each satellite's orbit is approximately $10^{-3}$. The stable constellation configuration of Taiji allows it to effectively detect gravitational waves within a sensitive frequency range spanning from $10^{-4}$ to 0.1 Hz~\cite{luo2021taiji, Ren:2023yec}.

One of Taiji's objectives is the detection of extreme-mass-ratio inspiral systems~\cite{Xin:2018urr, Han:2018hby, Yang:2019xro, Ren:2023yec}. These systems consist of a stellar-mass black hole orbiting around a much heavier black hole with a mass ranging from $10^4-10^7M_{\odot}$.  These supermassive black holes are typically located at the center of a galaxy~\cite{lynden1971quasars, soltan1982masses, Kormendy:1995er, Genzel:2010zy, Babak:2017tow}. When compact objects, such as black holes, orbit the  central black hole in EMRI systems~\cite{Volonteri:2009vh}, they can release energy and 
generate gravitational wave (GW) at low frequencies. The frequencies of these GWs fall within the sensitive range of space detectors~\cite{Amaro-Seoane:2012aqc, Amaro-Seoane:2012vvq}. The exploration of EMRI systems presents opportunities for testing the principles of general relativity~\cite{Babak:2017tow, Han:2018hby, Xin:2018urr, Shen:2023pje}. It can also  map the evolution of center massive black holes (MBH) by the reference of sources' parameters. These researches contribute to our understanding of the distribution of MBH masses and their relationships with host galaxies~\cite{Destounis:2022obl}. EMRI systems also provide a precise means to study and map the gravitational field of the central black hole~\cite{Maselli:2021men}. 

Analyzing EMRI signals  poses a substantial challenge due to their prolonged duration and intricate characteristics. These signals can endure for extended periods, spanning from several months to years, demanding significant computational resources  for waveform generation for comprehensive analysis~\cite{Cornish:2008zd, Babak:2009ua}. Traditional GW data analysis techniques, such as match filtering, would necessitate a minimum of $10^{40}$ templates to process a single waveform~\cite{Gair:2004iv}. Even with the application of the F–statistic algorithm~\cite{Wang:2012xh, Wang:2015kja} can reduce template requirements and enhance search and analysis efficiency, a substantial amount of time is still required, and there is a heavy reliance on templates.  EMRI signals  are inherently complicated, characterized by their intricate composition of harmonics and modulations. Additionally, the influence of self-force effects further compounds the difficulty of creating accurate models for these intricate signals. Although recent research has made progress in addressing the challenges posed by self-force~\cite{Lynch:2021ogr, Isoyama:2021jjd, Shen:2023pje}, it's vital to recognize the lack of precise and quickly generated EMRI waveform templates will introduce an additional layer of complexity into the data analysis process. 

Deep learning techniques, particularly Convolutional Neural Networks (CNNs), which has achieved great success in computer vision, have found substantial utility in the realm of GW astrophysics to detect and characterize GW signals originating from BBH mergers~\cite{George:2017pmj, George:2016hay, Wei:2019zlc, Wang:2019zaj}. These investigations have highlighted the effectiveness of deep learning in detecting both simulated and real GW signals  in the detectors' noise.  The application of deep learning has been expanded to the detection  of diverse types of gravitational wave (GW) sources.  In a recent study Ref.~\cite{Jin:2023ahl}, the UNet architecture was employed  to detect BBHs and BNSs by analyzing Q-transformed representations for GW data. Furthermore, a novel approach known as the matched-filtering CNN (MFCNN) was developed in Ref.~\cite{Ruan:2021fxq} to identify the mergers of Massive Binary Black Holes (MBHBs). This innovative method combines the strengths of matched filtering and CNN, resulting in a significant enhancement in the efficiency of identifying GW candidates compared to using matched filtering in isolation.  Furthermore, deep learning has demonstrated its effectiveness in GW data analysis for other space-based GW sources especially for Extreme Mass EMRIs~\cite{Zhang:2022xuq, Zhao:2022qob, Zhao:2023ncy}.

In our current research, we employ a straightforward 2-layer CNN network specifically designed for the detection of EMRI signals within time-frequency data. Our data preparation involves applying the Q-transform to convert the time series into time-frequency data before feeding it into the network. Previous research~\cite{Gair:2007bz, Gair:2008ec} have emphasized the utility of the time-frequency algorithm in analyzing EMRI signals.  They successfully detected and retrieved signals in the LISA Data Challenge. Additionally, Ref.~\cite{Jin:2023ahl} has highlighted the effectiveness of frequency-time data in the deep learning application. To enhance computational efficiency, we employ time-frequency data in our analysis and adjust it to smaller size. Our model efficiently processes 0.5-year Taiji data within seconds. We generate the signal we aim to detect by the AKK waveform model and simulate noise corresponding Taiji's analytical Power Spectral Density (PSD). Additionally, we incorporate second-generation Time Delay Interferometry (TDI)~\cite{Amaro-Seoane:2012vvq} to improve the model's applicability in practical scenarios. While our study primarily focuses on Taiji data, our model's versatility allows for easy adaptation to other space-based GW detectors like LISA.

The structure of the paper is as follows: Sec.~II serves as an introduction, delving into the characteristics of EMRI and its waveform models, the orbit of the Taiji detector, the Power Spectral Density , and the calculation of TDI. In Sec.~III,  we provide a comprehensive description of how we prepare and preprocess our training data. We also delve into the CNN architecture and present the outcomes of our testing. Lastly, in Sec.~IV, we provide a concise summary of our conclusions.

\section{EMRI Detection in space-based detectors}

When we analyze data from Taiji detector, it is crucial to create models specifically for the EMRI signals and for the  noise. In the following section, we will explain the fundamental concept of the simulation process, delve into the specifics of using Time Delay Interferometry (TDI) for processing Taiji detectors' data during simulations, and outline the approach we use to generate EMRI signals.

\subsection{Equal arm analytic orbit}
The Taiji detector is a space-based observatory that consists of three spacecrafts arranged in a triangular configuration. It uses lasers to measure distances between these spacecrafts, allowing it to detect gravitational waves in a frequency range of $10^{-4}$~Hz to $10^{-1}$~Hz.   
Currently, we utilize a simplified Keplerian orbit model to approximate the trajectories of the Taiji spacecrafts (SCs). It's essential to note that these orbits are basic and do not capture the full complexity of actual orbital dynamics. The depiction of the orbital motion for each Taiji spacecraft (SCn, where n = 1, 2, 3) is outlined below:
\begin{equation}
\begin{aligned}
& x_n=a \cos \alpha+a e\left(\sin \alpha \cos \alpha \sin \beta_n-\left(1+\sin ^2 \alpha\right) \cos \beta_n\right) \\
& y_n=a \sin \alpha+a e\left(\sin \alpha \cos \alpha \cos \beta_n-\left(1+\cos ^2 \alpha\right) \sin \beta_n\right) \\
& z_n=-\sqrt{3} a e \cos \left(\alpha-\beta_n\right)
\end{aligned}
\end{equation}~\cite{babak2020lisa}. 
The coordinates are defined within the solar system barycentric (SSB) frame. The angles $\beta_n$ are given by $(n-1) \frac{2 \pi}{3}+\lambda$, and $\alpha(t)$ is defined as $\frac{2 \pi}{1 \text { year }} t+\kappa$.

Initially, both $\lambda$ and $\kappa$ are set to zero. `a' is 1 astronomical unit (AU), and the orbital eccentricity  `e' is calculated as $e=L /(2 a \sqrt{3})$, where `L' corresponds to the distance between two spacecrafts. The distance designed for the Taiji mission is  3 kilometers~\cite{Hu:2017mde}.

\subsection{The  PSD and noise of Taiji}

In the practical context of Taiji detection, we must take into account various factors that contribute to noise. To streamline our simulation process, we've opted for a simplified noise model consisting of two distinct components. The first component, denoted as $P_{\mathrm{oms}}(f)$, represents high-frequency noise originating from the optical metrology system. The second component addresses low-frequency noise, $P_{\mathrm{acc}}(f)$, resulting from the test mass's acceleration. This noise can be effectively characterized through its power spectral density.

The PSD of the X channel for second-generation TDI can be mathematically expressed as follows~\cite{Ren:2023yec}:
\begin{equation}
\begin{gathered}
\mathrm{PSD}{X_2} = 64 \sin^2(\omega L) \sin^2(2 \omega L) \left(P{\mathrm{oms}} + (3 + \cos(2 \omega L)) P_{\mathrm{acc}}\right),
\end{gathered}
\end{equation}
where $\omega = 2 \pi f / c$. The functions $P_{\mathrm{oms}}(f)$ and $P_{\mathrm{acc}}(f)$ can be described as follows:
 \begin{equation}
\begin{aligned}
P_{\mathrm{oms}}(f) & =64 \times 10^{-24} \frac{1}{\mathrm{~Hz}}\left[1+\left(\frac{2 \mathrm{mHz}}{f}\right)^4\right]\left(\frac{2 \pi f}{c}\right)^2 \\
P_{\mathrm{acc}}(f) & =9 \times 10^{-30} \frac{1}{\mathrm{~Hz}}\left[1+\left(\frac{0.4 \mathrm{mHz}}{f}\right)^2\right]\left[1+\left(\frac{f}{8 \mathrm{mHz}}\right)^4\right]\left(\frac{1}{2 \pi f c}\right)^2
\end{aligned}
\end{equation}

\subsection{Time-delay interferometry}
In missions for a space-based detector, Time Delay Interferometry (TDI) plays a crucial role, serving as a vital technique to efficiently  suppress laser frequency noise and attain the desired sensitivity levels. TDI's fundamental principle involves precise time shifting and combination of measurements to create an interferometry with  equal-arm configuration. First-generation TDI combinations excel in mitigating laser frequency noise in static unequal-arm configurations, while second-generation TDI combinations extend this noise reduction capability to scenarios involving relative motion. In this paper, we exclusively employ the second-generation TDI configuration. The Michelson combination  in our paper is~\cite{Tinto:2003vj, Ren:2023yec}:
\begin{equation}
\begin{aligned}
& X_2(t)=y_{1^{\prime}}+y_{3,2^{\prime}}+y_{1,22^{\prime}}+y_{2^{\prime}, 322^{\prime}}+y_{1,3^{\prime} 322^{\prime}}+y_{2^{\prime}, 33^{\prime} 322^{\prime}} \\
& \quad+y_{1^{\prime}, 3^{\prime} 33^{\prime} 322^{\prime}}+y_{3,2^{\prime} 3^{\prime} 33^{\prime} 322^{\prime}}-y_1-y_{2^{\prime}, 3}-y_{1^{\prime}, 3^{\prime} 3}-y_{3 ; 2^{\prime} 3^{\prime} 3} \\
& \quad-y_{1^{\prime} .22^{\prime} 3^{\prime} 3}-y_{3.2^{\prime} 22^{\prime} 3^{\prime} 3}-y_{1: 22^{\prime} 22^{\prime} 3^{\prime} 3}-y_{2^{\prime}, 322^{\prime} 22^{\prime} 3^{\prime} 3}
\end{aligned}
\end{equation}

The variables Y and Z are obtained through cyclic permutation of the indices. An uncorrelated set of TDI variables, denoted as A, E, and T, can be derived from linear combinations of X, Y, and Z, as described by ~\cite{Vallisneri:2004bn}:
\begin{equation}
\begin{aligned}
& A = \frac{1}{\sqrt{2}}(Z-X), \
& E = \frac{1}{\sqrt{6}}(X-2Y+Z), \
& T = \frac{1}{\sqrt{3}}(X+Y+Z).
\end{aligned}
\end{equation}
In this paper, all simulation data are processed through TDI, and we will specifically use the A and E channel data for analysis. We use \textbf{fastlisaresponse}(https://github.com/mikekatz04/lisa-on-gpu/tree/master) to generate the data after TDI in this work~\cite{Katz:2022yqe}.

\subsection{EMRI waveform model}

\begin{figure*}
\includegraphics[width=0.6\textwidth]{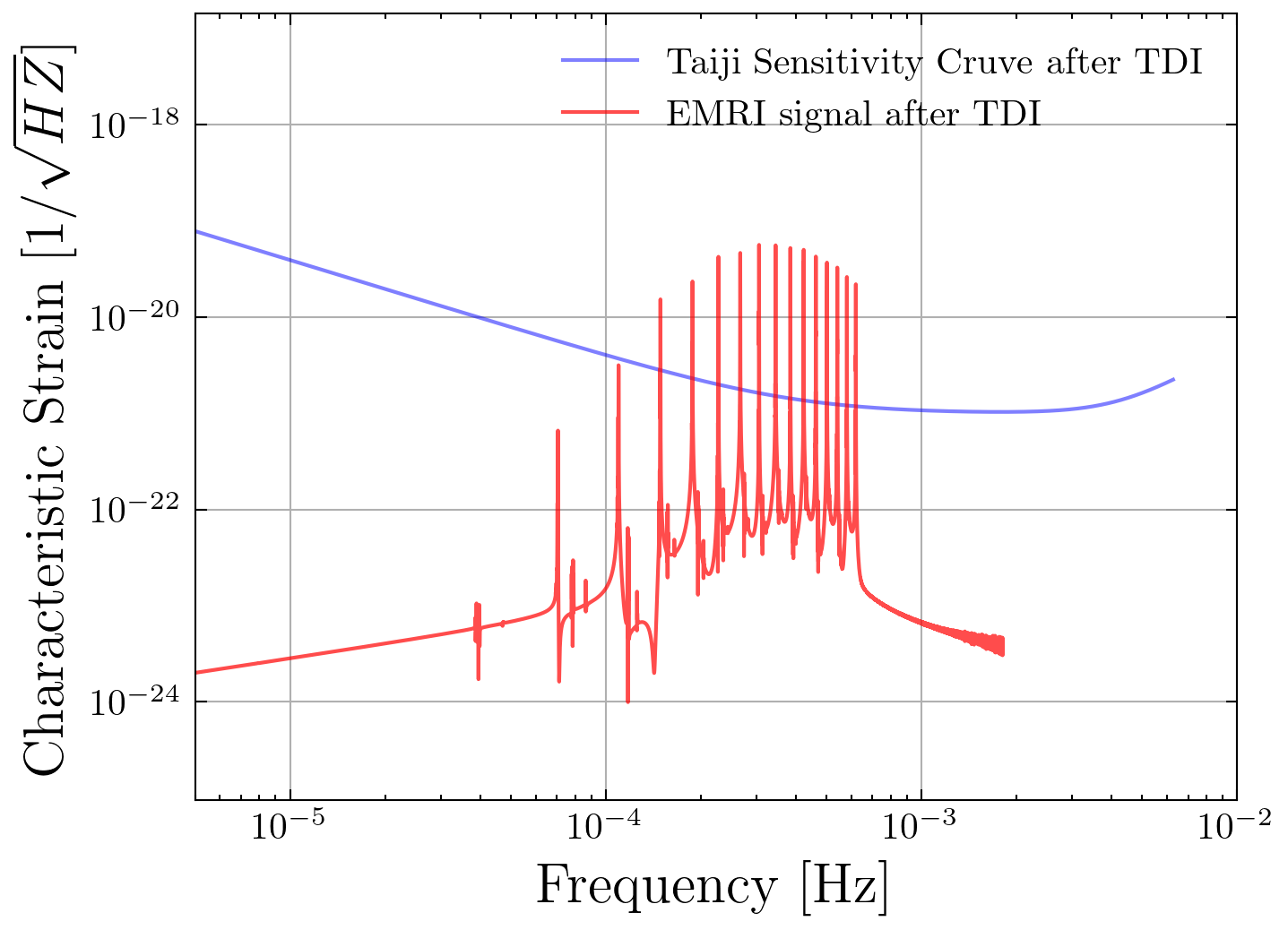} 

\caption{ A EMRI signal with SNR=50 and the Taiji's sensitivity curve after TDI, assuming the observation time is 0.5 year. The SNR value is 50.}
\label{EMRI_fd}
\end{figure*}

\begin{figure*}
\includegraphics[width=1\textwidth]{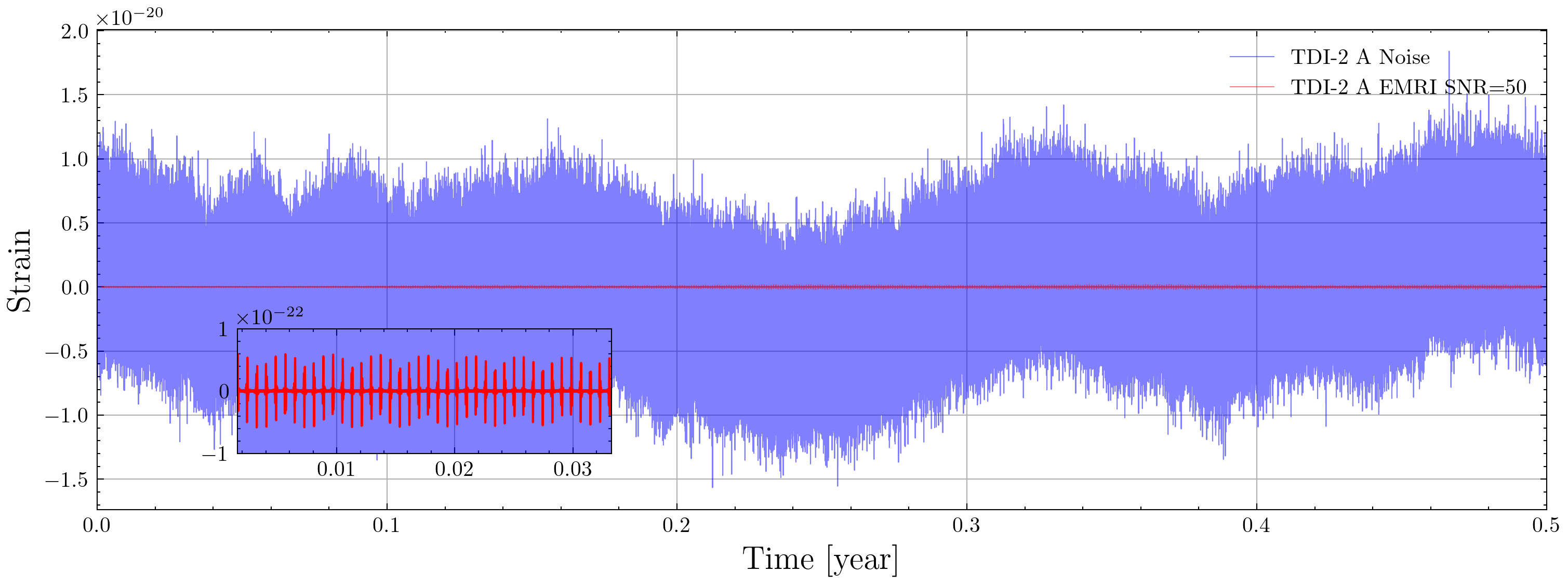} 

\caption{An example of an EMRI signal and noise in the A channel of the Taiji detector, comparing it to the EMRI signal hI in the A channel alone. All data has been processed through TDI. In this scenario, the signal-to-noise ratio  is configured to be 50, and the observation duration is 0.5 years.}
\label{EMRI_td}

\end{figure*}

The Taiji gravitational wave detector possesses the capability to identify EMRI events involving a compact object~(CO) and a massive black hole, covering a mass range from $10^4 M\odot$ to $10^7 M\odot$. The probability of detection increases as the system achieves a signal-to-noise ratio  exceeding 20.  Before the detection of EMRI signals, it is essential to have theoretical waveform models for EMRI.

In the analysis of EMRI events, researchers require rapid and precise waveform models. These models should be able to handle  sources with various parameters.  The AK model~\cite{Barack:2003fp}, while computationally efficient, simplifies gravitational wave radiation and provides only an estimation of strong field dynamics, deviating from the genuine signal. To address this, Ref~\cite{Chua:2017ujo} explored  the AKK model, which extends fading effects until the frequency reaches a stable Kerr orbit~\cite{Babak:2006uv}.  The AAK waveform strikes a balance between accuracy and computational efficiency, enhancing our ability to model EMRI events effectively. 

Assuming the spin of the compact object is negligible, an EMRI can be described by 14 parameters. 7 is the intrinsic parameters :$\left(\mu, M, a, e_0, \iota_0, \gamma_0, \psi_0\right)$. $M$ is the mass of the massive black hole, $\mu$ is the mass of the CO.  `a' is spin parameter of the MBH, `p' is the semi-latus rectum, `e' is eccentricity, $ \iota$ is the orbit’s inclination angle from the equatorial plane. $\gamma:=\tilde{\gamma}-\beta$ and $\beta(\hat{\mathbf{R}}, \hat{\mathbf{S}}, \hat{\mathbf{L}})=\beta\left(\theta_S, \phi_S, \theta_K, \phi_K, \iota, \alpha\right)$ is an azimuthal angle in the orbital plane. True anomaly $\psi$ is the angle that describes an object's position in an elliptical orbit relative to the central body and the periapsis. Rest parameters are the extrinsic parameters: $\left(p_0, \theta_S, \phi_S, \theta_K, \phi_K, \alpha_0, D\right)$.
`p' is the semi-latus rectum,  $\theta_S$  and $\theta_K$ are the polar and azimuthal sky location angles. \
$\theta_K$ and $\phi_K$ are the azimuthal and polar angles describing the orientation of the spin angular momentum vector of the MBH. 
The time dependence of the orbital orientation is confined to $\alpha~(t)$, $D$ is the luminosity distance~\cite{Chua:2017ujo}. The EMRI signals in the training data for this study have all been generated using the AAK method through the \textbf{EMRI\_Kludge\_Suite}~\cite{Chua:2017ujo}.

\section{ SEARCH STRATEGY}
\subsection{Datasets}

In order to employ a Convolutional Neural Network  for detecting EMRI signals within the Taiji detector data, we start by dividing the dataset into two groups, each comprising half of the samples. One of these sets contains data encompassing both signals and noise ($d = h + n$) and is assigned the label 1, whereas the other set exclusively contains noise ($n$) and is assigned the label 0.  Subsequently, the dataset is further divided into two sets: training data and testing data. The training data is employed for model training, while the testing data serves as an evaluation  for the trained CNN.

We make use of the AAK model to create EMRI signals, with the range of the parameters specified in Table~\ref{paramters}. The signals extend for a duration of 0.5 years and  are sampled every 10 seconds ($dt$). Following this, we apply the analytical Power Spectral Density to simulate noise, ensuring it aligns with the generated signal, which leads to the generation of both datasets $d$ and $n$.The total number of data generated for both $d$ and $n$ amounts to 6000.

\begin{table}[ht]
\centering
\renewcommand{\arraystretch}{1.6}
\caption{The parameters range of AKK waveforms for training and testing data}
\begin{tabular}{p{3.5cm}|p{4.5cm}}
\hline\hline
Parameter & Range (Uniform distribution)\\
\hline
$M/M_{\odot}$ & ($10^5, 10^8$) \\
$\mu/M_{\odot}$ & (10, 100) ; $M/\mu \geq 10^4 $\\
$a$ & $(10^{-3}$, 0.90) \\
$e_0$ & (0.005, 0.5) \\
$p_0/M$ & (10, 12) \\
$\iota$ & (0, 0.1) \\
$\gamma$ & (0, 0.1) \\
$\theta_S$ & (0, $\pi$) \\
$\phi_S$ & (0, $2\pi$ )\\
$\theta_K$ & (0, $\pi$) \\
$\phi_K$ & (0, $2\pi$) \\
\hline\hline
\end{tabular}
\label{paramters}
\end{table}
Subsequently, we incorporate Time Delay Interferometry (TDI) into the dataset, getting the A, E, and T channel data. For our training and testing purposes, we select data from the A and E channels. The datasize is (2,1560000) for both $d$ and $n$ . After TDI, we proceed to compute the Signal-to-Noise Ratio values, carefully maintaining them in the certain range.

We define the characteristic strain $h_{\mathfrak{c}}$ and noise amplitude $h_{n}$ as follows~\cite{Moore:2014lga}:
\begin{equation}
    \begin{aligned}&\left[h_{\text{c}}(f)\right]^{2}=4f^{2}\Big|\tilde{h}(f)\Big|^{2},\\\\&\left[h_{n}(f)\right]^{2}=f\text{S}_{n}(f),\end{aligned}
\end{equation}

So that we can express the SNR in equation~\ref{eq:snr}. We utilize $h_{\mathfrak{c}}(f)$ and $h(f)$ in the SNR calculation. When these values are visualized using a logarithmic scale, the area between the source and detector curves represents the SNR. This SNR value indicates the detectability of the source. We consider an EMRI signal to be a ``golden" EMRI when its SNR exceeds 50. Consequently, we adjust the distance to ensure that the SNR remains within the range of (50, 100). We use $\varrho$ to represent the SNR value in each channel:
\begin{equation}
  \varrho^2=\int_{-\infty}^\infty\mathrm{d}(\log f)\left[\frac{h_\mathrm{c}(f)}{h_n(f)}\right]^2  
  \label{eq:snr}
\end{equation}
When taking two channels into account, we get the target SNR using the following equation:
\begin{equation}
    \mathrm{SNR}^2=(Q_A)^2+(Q_E)^2.
\end{equation}
\subsection{The Q-transform}

Time-frequency analysis has proven to be highly effective in the EMRI data analysis. In Ref.~\cite{Wang:2012xh}, a time-frequency plot depicting a typical EMRI signal, wherein distinct frequency tracks clearly reveal the presence of dominant harmonics. The EMRI signals show a marked concentration of frequency components within these tracks, displaying substantial intensity.
 Therefore, we have chosen to employ the Constant Q Transform (CQT) for data preprocessing in this deep learning project. CQT is the preferred choice due to its advantages in this particular endeavor.

 In signal processing, time-frequency analysis encompasses techniques that examine a signal in both the time and frequency domains simultaneously. The use of time-frequency analysis offers several key advantages. Firstly, it allows for the adjustment of the trade-off between time and frequency resolution, providing greater flexibility in resolution settings. Additionally, time-frequency representations can extract relevant features crucial for deep learning and classification tasks. In comparison to time series analysis, time-frequency analysis significantly enhances feature extraction capabilities, making it a valuable tool. Time-frequency analysis techniques encompass a variety of methods such as Short-Time Fourier Transform (STFT), Wavelet Transform, Continuous Wavelet Transform (CWT), and more.

In our research, we have decided to employ the Constant Q Transform (CQT) for data processing in our deep learning project. We have chosen CQT over the Short Time Fourier Transform (STFT) for several reasons. Firstly, CQT offers adaptable time and frequency resolution, unlike STFT, which has a fixed resolution determined by window size and window overlap. Generally, CQT provides finer frequency details for higher frequencies and vice versa. Secondly, CQT's flexible time and frequency resolution, along with its exceptional ability to accurately pinpoint both time and frequency information, makes it the preferred choice for visualizing and analyzing signals like EMRI, especially when compared to STFT.

CQT is a special case of Variable Q transform(VQT). Both of them are related to complex Morlet wavelet transform. The definition is
\begin{equation}
    \delta f_k=2^{1/n}\cdot\delta f_{k-1}=\left(2^{1/n}\right)^k\cdot\delta f_{\min},
\end{equation}

where $\delta f_k$ is the bandwidth of the k-th filter, $f_{min}$ is the central frequency of the lowest filter, and n is the number of filters per octave  (https://en.wikipedia.org/wiki/Constant-Q\_transform).

Define the quality factor Q for  CQT:
\begin{equation}
    Q=\frac{f_k}{\delta f_k}.
\end{equation}

With this factor, we can define window length: 
\begin{equation}
    N[k]=\frac{f_{s}}{\delta f_{k}}=\frac{f_{s}}{f_{k}}Q.
\end{equation}

Hamming window is one of the most commonly used window, take this for example:
\begin{equation}
    W[k,n]=\alpha-(1-\alpha)\cos\frac{2\pi n}{N[k]-1},\quad\alpha=25/46,\quad0\leqslant n\leqslant N[k]-1.
\end{equation}

With things above, define Q transform as:
\begin{equation}
    X[k]=\frac{1}{N[k]}\sum_{n=0}^{N[k]-1}W[k,n]x[n]e^{\frac{-j2\pi Qn}{N[k]}}.
\end{equation}

\begin{figure*}
\subfigure[\label{fig:puresignal}]{
\includegraphics[width=0.45\textwidth]{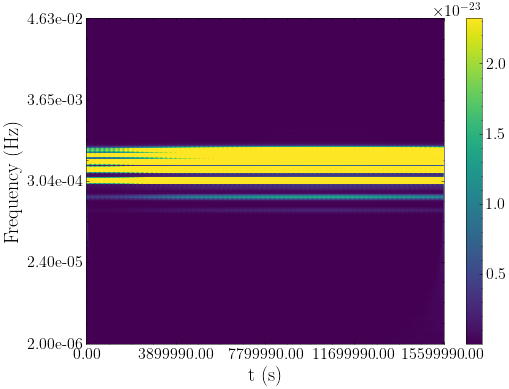}}
\subfigure[\label{fig:signal_snr100}]{
\includegraphics[width=0.45\textwidth]{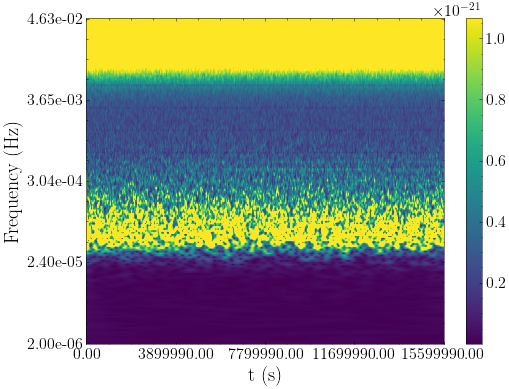}}
\caption{Time-frequency domain EMRI signals after Q-transform with the SNR=100 and along with the signal embedding in the noise. }
\label{fig:noise_rgb}
\end{figure*}

\begin{figure}
\centering
\subfigure[\label{fig:purenoise}]{
\includegraphics[width=0.45\textwidth]{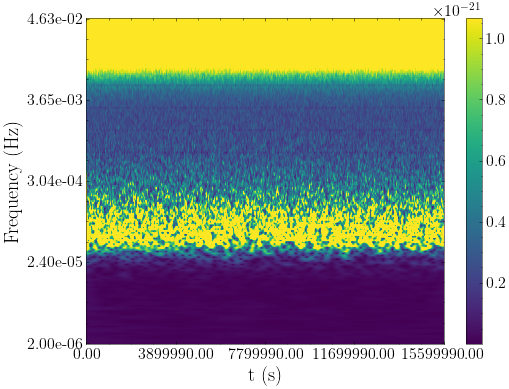}}
\subfigure[\label{fig:signal_snr50}]{
\includegraphics[width=0.45\textwidth]{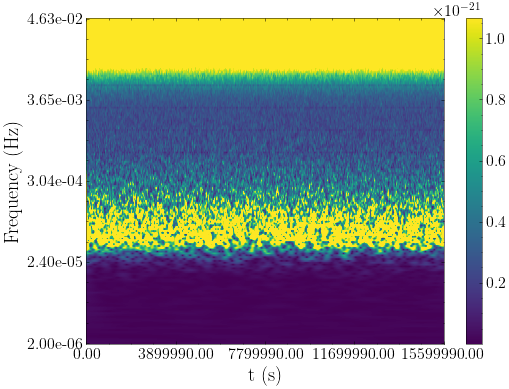}}
\caption{Time-frequency domain noise and EMRI signals embedding in the detector's noise with a SNR = 50 after Q-transform.}
\label{fig:signal}
\end{figure}

We ues \textbf{librosa}(https://librosa.org/doc/latest/index.html) to do the process. In Fig.~\ref{fig:puresignal}, we see the signal after the Q-transform. The EMRI signal is strong, with a high SNR of 100. In Figure~\ref{fig:signal_snr100}, we can observe the signal amidst the background noise of the detector. It's noticeable that only a few of its  features remain preserved after adding noise. In Figure~\ref{fig:signal}, we can see the pure noise denoted as $n$, and  the data $d$ that includes both the signal and the noise after undergoing the Q-transform. In Figure~\ref{fig:signal_snr50}, where the signal-to-noise ratio is 50, the distinctive features of the EMRI are obscured by the noise, making it difficult to distinguish between Figure~\ref{fig:purenoise} and Figure~\ref{fig:signal_snr50} by visual inspection alone. However, employing convolutional neural networks (CNN) can make this differentiation quite easy.

We store the Q-transform results as RGB images and resize them to (255,255) matrixs to use them as input for the CNN network.

\subsection{Network}
\begin{center}
\begin{table}
\centering
\renewcommand{\arraystretch}{1.6} 
\caption{ Architecture of deep convolutional neural network}
\label{network}
\begin{tabular}{|p{1cm}|p{8cm}|p{4.0cm}|p{2.2cm}|p{2cm}|}
\hline
\textbf{num} & \textbf{Layer (type)} & \textbf{Output Shape} & \textbf{Param } \\
\hline\hline
1&Conv2d (Kernel Size: 3x3 Stride: 1 Padding: 1 )& [16, 16, 225, 225] & 448 \\
2&BatchNorm2d & [16, 16, 225, 225] & 32 \\
3&GELU & [16, 16, 225, 225] & 0 \\
4&MaxPool2d (Kernel Size: 2x2 Stride: 2) & [16, 16, 112, 112] & 0 \\
5&Conv2d (Kernel Size: 3x3 Stride: 1 Padding: 1)& [16, 32, 112, 112] & 4,640 \\
6&BatchNorm2d & [16, 32, 112, 112] & 64 \\
7&GELU & [16, 32, 112, 112] & 0 \\
8&MaxPool2d (Kernel Size: 2x2 Stride: 2)& [16, 32, 56, 56] & 0 \\
9&Flatten & [16, 100352] & 0 \\
10&Linear & [16, 128] & 12,845,184 \\
11&BatchNorm1d-11 & [16, 128] & 256 \\
12&GELU & [16, 128] & 0 \\
13&Dropout & [16, 128] & 0 \\
14&Linear & [16, 64] & 8,256 \\
15&Dropout & [16, 64] & 0 \\
16&BatchNorm1d & [16, 64] & 128 \\
17&GELU & [16, 64] & 0 \\
18&Linear & [16, 2] & 130 \\
\hline\hline

\end{tabular}
\end{table}
\end{center}
This study adopts the basic framework of a CNN network architecture, which includes Conv2d layers with filter sizes of 16 and 32, followed by two fully connected layers with sizes of 128 and 64. Standard ReLU activation functions are employed as non-linearities between layers, and the convolutional layers have kernel sizes of $3 \times 3$ and $3 \times 3$. The network incorporates max-pooling layers. We uses a stride of 1 for convolution layers and a stride of 2 for pooling layers. Specific details about the network can be found in Table~\ref{network}.

In our approach, we utilize the cross-entropy loss function, which is a widely adopted choice in classification tasks due to its effectiveness in speeding up the convergence of deep learning models. This loss function quantifies the difference between predicted probabilities and actual labels, encouraging the model to improve its predictive accuracy. The mathematical expression for this loss function is as follows:
\begin{equation}
L = \sum_{j=1}^{N} \left( -y_{j_{\text{true}}} \cdot \log(y_{j_{\text{pred}}} ) - (1 - y_{j_{\text{true}}}) \cdot \log(1 - y_{j_{\text{pred}}}) \right)
\end{equation}

Specifically, for our binary classification task, which involves distinguishing EMRI signals from noise, the cross-entropy loss measures the discrepancy between the predicted probabilities of EMRI and the actual EMRI labels in the training data. By minimizing this loss, our model aims to align its predictions with the true EMRI labels, ultimately enhancing its ability to classify EMRI signals accurately and reliably.

Additionally, we implement the Adam optimizer with a learning rate equal to 0.01 to update the model's weights during training. Over the course of 200 training epochs for training data, we monitor both loss and accuracy, periodically saving the model with the lowest loss after the convergence as our best performing model. Then we test the saved model on test data.
\subsection{Result}

In this section, we use receiver operating characteristics (ROC)  to see how well our CNN model can find the signal in our testing data. The ROC curve is a visual tool that helps us see how well our model can find the signal in our data without messing up. The ROC curve serves as a graphical representation that provides insights into our model's performance. It focuses on two critical aspects: the true positive rate, indicating how frequently the model correctly identifies the signal, and the false positive rate, highlighting instances where the model incorrectly detects a signal when none exists. To provide an overarching assessment of our model's capabilities, we rely on the Area Under the ROC Curve (AUC). AUC is like a score which can helps us see how well our model is doing its job. A higher AUC value, closer to 1, signifies superior performance, indicating that the model excels in detecting the signal with precision.

\begin{figure*}
    
\includegraphics[width=0.45\textwidth]{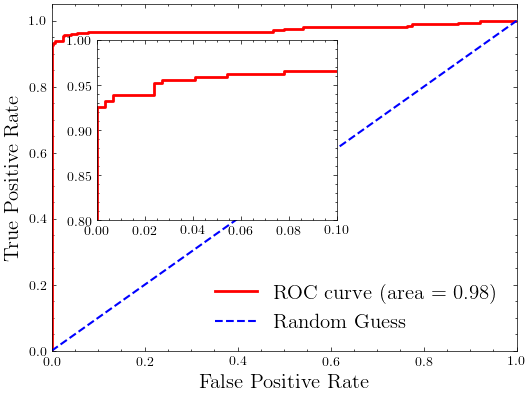}
\includegraphics[width=0.45\textwidth]{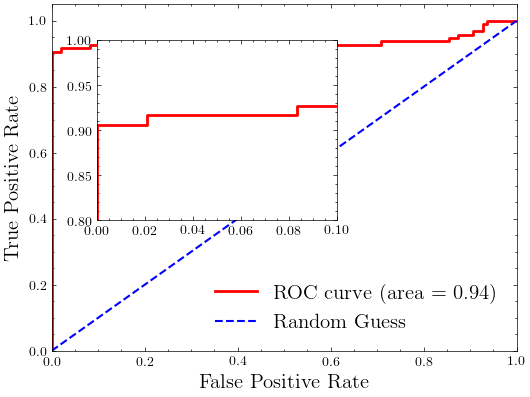}
\caption{ ROC analysis for the CNN model trained to detect EMRI signals. The SNRs of the testing datasets are selected to [50,100 ] and 50 only. The blue dotted line denotes the line of
random classifier. The small box indicates the region FPR at 0 to 0.1.}
\label{fig:roc}
\end{figure*}

In Figure~\ref{fig:roc}, there are two separate ROC curves, one in (a) and the other in (b). These curves represent different SNR values in the test data. In (a), the SNR values range from 50 to 100, while in (b), it's a constant value of 50. The AUC values shown in the figures give us a numerical measure of how good the model is at accurately spotting EMRI signals. To make it easier to see the detection performance, especially at low false positive rates, there's a smaller inset figure inside Fig.~\ref{fig:roc} that zooms in on the TPR from 0 to 0.01. This helps us understand how well our CNN network performs in detecting EMRI signals across different SNR values.

\section{Conclusion}
This study aims to address one of the most challenging issues in space-based gravitational wave detection: the identification of EMRIs. These signals are known for their complex and lengthy waveforms, as well as their inherently weak nature, making their detection quite challenging. To overcome this challenge, we present an innovative approach that combines the Q-transform for data preprocessing and a CNN network. This method not only preserves the crucial EMRI signal characteristics but also enhances efficiency, making it suitable for space-based detectors.

Our findings reveal the robustness of the CNN-based model, particularly in the detection of EMRI signals with a SNR of 50, often referred to as ``golden" EMRIs. Incorporating time-delay interferometry into our approach ensures its practical utility. The model demonstrates impressive performance, achieving a true positive rate of 94.2\% at a 1\% false positive rate  across a wide range of SNRs from 50 to 100. Even at an SNR of 50, the TPR remains high at 91\% while maintaining a low FPR of 1\%. This research opens new avenues for advancing EMRI data analysis through the integration of deep learning techniques with time-frequency data, holding great promise for enhancing the capabilities of space-based gravitational wave detectors.

In the future, successful implementation of this CNN-based approach opens the door to further improvements. Subsequent research endeavors could explore lower SNR thresholds, possibly as low as SNR 20, requiring the development of more extensive networks capable of detecting weak signals. Additionally, integrating more advanced simulations like self-force EMRI waveforms or complex noise models may demand increased computational resources. Traditional GW analysis methods may not suffice for such data analysis. As we start using machine learning to find EMRI signals, there's a lot we have not explored yet. This could lead to big advancements in finding space-based gravitational wave sources. Especially, if from machine learning, the preliminary range of source parameters can be set, then will be very powerful for the further parameter estimation.
\section{Acknowledgements*} 

This work was supported by the National Key R\&D Program of China (Grant Nos. 2021YFC2203002), the National Natural Science Foundation of China (Grant Nos. 12173071). Wen-Biao Han was supported by the CAS Project for Young Scientists in Basic Research (Grant No. YSBR-006). This work made use of the High Performance Computing Resource in the Core Facility for Advanced Research Computing at Shanghai Astronomical Observatory.


\begin{thebibliography}{99}

\bibitem{LIGOScientific:2016dsl}
B.~P. Abbott et~al.
\newblock {Binary Black Hole Mergers in the first Advanced LIGO Observing Run}.
\newblock {\em Phys. Rev. X}, 6(4):041015, 2016.
\newblock [Erratum: Phys.Rev.X 8, 039903 (2018)].

\bibitem{LIGOScientific:2014pky}
J.~Aasi et~al.
\newblock {Advanced LIGO}.
\newblock {\em Class. Quant. Grav.}, 32:074001, 2015.

\bibitem{LIGOScientific:2016aoc}
B.~P. Abbott et~al.
\newblock {Observation of Gravitational Waves from a Binary Black Hole Merger}.
\newblock {\em Phys. Rev. Lett.}, 116(6):061102, 2016.

\bibitem{LIGOScientific:2018urg}
B.~P. Abbott et~al.
\newblock {Search for gravitational waves from a long-lived remnant of the
  binary neutron star merger GW170817}.
\newblock {\em Astrophys. J.}, 875(2):160, 2019.

\bibitem{Harry:2010zz}
Gregory~M. Harry.
\newblock {Advanced LIGO: The next generation of gravitational wave detectors}.
\newblock {\em Class. Quant. Grav.}, 27:084006, 2010.

\bibitem{VIRGO:2014yos}
F.~Acernese et~al.
\newblock {Advanced Virgo: a second-generation interferometric gravitational
  wave detector}.
\newblock {\em Class. Quant. Grav.}, 32(2):024001, 2015.

\bibitem{LIGOScientific:2018mvr}
B.~P. Abbott et~al.
\newblock {GWTC-1: A Gravitational-Wave Transient Catalog of Compact Binary
  Mergers Observed by LIGO and Virgo during the First and Second Observing
  Runs}.
\newblock {\em Phys. Rev. X}, 9(3):031040, 2019.

\bibitem{LIGOScientific:2020ibl}
R.~Abbott et~al.
\newblock {GWTC-2: Compact Binary Coalescences Observed by LIGO and Virgo
  During the First Half of the Third Observing Run}.
\newblock {\em Phys. Rev. X}, 11:021053, 2021.

\bibitem{LIGOScientific:2021usb}
R.~Abbott et~al.
\newblock {GWTC-2.1: Deep Extended Catalog of Compact Binary Coalescences
  Observed by LIGO and Virgo During the First Half of the Third Observing Run}.
\newblock 8 2021.

\bibitem{LIGOScientific:2021djp}
R.~Abbott et~al.
\newblock {GWTC-3: Compact Binary Coalescences Observed by LIGO and Virgo
  During the Second Part of the Third Observing Run}.
\newblock 11 2021.

\bibitem{LIGOScientific:2020kqk}
R.~Abbott et~al.
\newblock {Population Properties of Compact Objects from the Second LIGO-Virgo
  Gravitational-Wave Transient Catalog}.
\newblock {\em Astrophys. J. Lett.}, 913(1):L7, 2021.

\bibitem{KAGRA:2021duu}
R.~Abbott et~al.
\newblock {Population of Merging Compact Binaries Inferred Using Gravitational
  Waves through GWTC-3}.
\newblock {\em Phys. Rev. X}, 13(1):011048, 2023.

\bibitem{Belczynski:2017gds}
K.~Belczynski et~al.
\newblock {Evolutionary roads leading to low effective spins, high black hole
  masses, and O1/O2 rates for LIGO/Virgo binary black holes}.
\newblock {\em Astron. Astrophys.}, 636:A104, 2020.

\bibitem{LIGOScientific:2019fpa}
B.~P. Abbott et~al.
\newblock {Tests of General Relativity with the Binary Black Hole Signals from
  the LIGO-Virgo Catalog GWTC-1}.
\newblock {\em Phys. Rev. D}, 100(10):104036, 2019.

\bibitem{LIGOScientific:2021sio}
R.~Abbott et~al.
\newblock {Tests of General Relativity with GWTC-3}.
\newblock 12 2021.

\bibitem{LIGOScientific:2021aug}
R.~Abbott et~al.
\newblock {Constraints on the Cosmic Expansion History from
  GWTC\textendash{}3}.
\newblock {\em Astrophys. J.}, 949(2):76, 2023.

\bibitem{Aso:2013eba}
Yoichi Aso, Yuta Michimura, Kentaro Somiya, Masaki Ando, Osamu Miyakawa,
  Takanori Sekiguchi, Daisuke Tatsumi, and Hiroaki Yamamoto.
\newblock {Interferometer design of the KAGRA gravitational wave detector}.
\newblock {\em Phys. Rev. D}, 88(4):043007, 2013.

\bibitem{KAGRA:2013rdx}
B.~P. Abbott et~al.
\newblock {Prospects for observing and localizing gravitational-wave transients
  with Advanced LIGO, Advanced Virgo and KAGRA}.
\newblock {\em Living Rev. Rel.}, 21(1):3, 2018.

\bibitem{Kaiser:2020tlg}
Andrew~R. Kaiser and Sean~T. McWilliams.
\newblock {Sensitivity of present and future detectors across the black-hole
  binary gravitational wave spectrum}.
\newblock {\em Class. Quant. Grav.}, 38(5):055009, 2021.

\bibitem{LISA:2017pwj}
Pau Amaro-Seoane et~al.
\newblock {Laser Interferometer Space Antenna}.
\newblock 2 2017.

\bibitem{Ni:2012eh}
Wei-Tou Ni.
\newblock {ASTROD-GW: Overview and Progress}.
\newblock {\em Int. J. Mod. Phys. D}, 22:1341004, 2013.

\bibitem{Kawamura:2011zz}
Seiji Kawamura et~al.
\newblock {The Japanese space gravitational wave antenna: DECIGO}.
\newblock {\em Class. Quant. Grav.}, 28:094011, 2011.

\bibitem{Crowder:2005nr}
Jeff Crowder and Neil~J. Cornish.
\newblock {Beyond LISA: Exploring future gravitational wave missions}.
\newblock {\em Phys. Rev. D}, 72:083005, 2005.

\bibitem{Hogan:2011tsw}
Jason~M. Hogan et~al.
\newblock {An Atomic Gravitational Wave Interferometric Sensor in Low Earth
  Orbit (AGIS-LEO)}.
\newblock {\em Gen. Rel. Grav.}, 43:1953--2009, 2011.

\bibitem{Hu:2017mde}
Wen-Rui Hu and Yue-Liang Wu.
\newblock {The Taiji Program in Space for gravitational wave physics and the
  nature of gravity}.
\newblock {\em Natl. Sci. Rev.}, 4(5):685--686, 2017.

\bibitem{Ruan:2018tsw}
Wen-Hong Ruan, Zong-Kuan Guo, Rong-Gen Cai, and Yuan-Zhong Zhang.
\newblock {Taiji program: Gravitational-wave sources}.
\newblock {\em Int. J. Mod. Phys. A}, 35(17):2050075, 2020.

\bibitem{TianQin:2015yph}
Jun Luo et~al.
\newblock {TianQin: a space-borne gravitational wave detector}.
\newblock {\em Class. Quant. Grav.}, 33(3):035010, 2016.

\bibitem{Klein:2015hvg}
Antoine Klein et~al.
\newblock {Science with the space-based interferometer eLISA: Supermassive
  black hole binaries}.
\newblock {\em Phys. Rev. D}, 93(2):024003, 2016.

\bibitem{Babak:2017tow}
Stanislav Babak, Jonathan Gair, Alberto Sesana, Enrico Barausse, Carlos~F.
  Sopuerta, Christopher P.~L. Berry, Emanuele Berti, Pau Amaro-Seoane, Antoine
  Petiteau, and Antoine Klein.
\newblock {Science with the space-based interferometer LISA. V: Extreme
  mass-ratio inspirals}.
\newblock {\em Phys. Rev. D}, 95(10):103012, 2017.

\bibitem{Han:2018hby}
Wen-Biao Han and Xian Chen.
\newblock {Testing general relativity using binary extreme-mass-ratio
  inspirals}.
\newblock {\em Mon. Not. Roy. Astron. Soc.}, 485(1):L29--L33, 2019.

\bibitem{Xin:2018urr}
Shuo Xin, Wen-Biao Han, and Shu-Cheng Yang.
\newblock {Gravitational waves from extreme-mass-ratio inspirals using general
  parametrized metrics}.
\newblock {\em Phys. Rev. D}, 100(8):084055, 2019.

\bibitem{Yang:2019xro}
Shucheng Yang, Shuo Xin, Chen Zhang, and Wenbiao Han.
\newblock {Testing Gravity Theory With Extreme Mass-Ratio Inspirals: Recent
  Progress}.
\newblock {\em MDPI Proc.}, 17(1):11, 2019.

\bibitem{Lamberts:2019nyk}
Astrid Lamberts, Sarah Blunt, Tyson~B. Littenberg, Shea Garrison-Kimmel, Thomas
  Kupfer, and Robyn~E. Sanderson.
\newblock {Predicting the LISA white dwarf binary population in the Milky Way
  with cosmological simulations}.
\newblock {\em Mon. Not. Roy. Astron. Soc.}, 490(4):5888--5903, 2019.

\bibitem{Caprini:2015zlo}
Chiara Caprini et~al.
\newblock {Science with the space-based interferometer eLISA. II: Gravitational
  waves from cosmological phase transitions}.
\newblock {\em JCAP}, 04:001, 2016.

\bibitem{Bartolo:2016ami}
Nicola Bartolo et~al.
\newblock {Science with the space-based interferometer LISA. IV: Probing
  inflation with gravitational waves}.
\newblock {\em JCAP}, 12:026, 2016.

\bibitem{Gong:2011zzd}
Xue-Fei Gong et~al.
\newblock {A scientific case study of an advanced LISA mission}.
\newblock {\em Class. Quant. Grav.}, 28:094012, 2011.

\bibitem{luo2021taiji}
Ziren Luo, Yan Wang, Yueliang Wu, Wenrui Hu, and Gang Jin.
\newblock The taiji program: A concise overview.
\newblock {\em Progress of Theoretical and Experimental Physics},
  2021(5):05A108, 2021.

\bibitem{Ren:2023yec}
Zhixiang Ren, Tianyu Zhao, Zhoujian Cao, Zong-Kuan Guo, Wen-Biao Han, Hong-Bo
  Jin, and Yue-Liang Wu.
\newblock {Taiji data challenge for exploring gravitational wave universe}.
\newblock {\em Front. Phys. (Beijing)}, 18(6):64302, 2023.

\bibitem{lynden1971quasars}
Donald Lynden-Bell and Martin~J Rees.
\newblock On quasars, dust and the galactic centre.
\newblock {\em Monthly Notices of the Royal Astronomical Society},
  152(4):461--475, 1971.

\bibitem{soltan1982masses}
Andrzej Soltan.
\newblock Masses of quasars.
\newblock {\em Monthly Notices of the Royal Astronomical Society},
  200(1):115--122, 1982.

\bibitem{Kormendy:1995er}
John Kormendy and Douglas Richstone.
\newblock {Inward bound: The Search for supermassive black holes in galactic
  nuclei}.
\newblock {\em Ann. Rev. Astron. Astrophys.}, 33:581, 1995.

\bibitem{Genzel:2010zy}
Reinhard Genzel, Frank Eisenhauer, and Stefan Gillessen.
\newblock {The Galactic Center Massive Black Hole and Nuclear Star Cluster}.
\newblock {\em Rev. Mod. Phys.}, 82:3121--3195, 2010.

\bibitem{Volonteri:2009vh}
Marta Volonteri and Priyamvada Natarajan.
\newblock {Journey to the $M_{\rm BH} - \sigma$ relation: the fate of low mass
  black holes in the Universe}.
\newblock {\em Mon. Not. Roy. Astron. Soc.}, 400:1911, 2009.

\bibitem{Amaro-Seoane:2012aqc}
Pau Amaro-Seoane et~al.
\newblock {eLISA/NGO: Astrophysics and cosmology in the gravitational-wave
  millihertz regime}.
\newblock {\em GW Notes}, 6:4--110, 2013.

\bibitem{Amaro-Seoane:2012vvq}
Pau Amaro-Seoane et~al.
\newblock {Low-frequency gravitational-wave science with eLISA/NGO}.
\newblock {\em Class. Quant. Grav.}, 29:124016, 2012.

\bibitem{Shen:2023pje}
Ping Shen, Wen-Biao Han, Chen Zhang, Shu-Cheng Yang, Xing-Yu Zhong, and
  Ye~Jiang.
\newblock {The influence of mass-ratio in extreme-mass-ratio inspirals for
  testing general relativity}.
\newblock 3 2023.

\bibitem{Destounis:2022obl}
Kyriakos Destounis, Arun Kulathingal, Kostas~D. Kokkotas, and Georgios~O.
  Papadopoulos.
\newblock {Gravitational-wave imprints of compact and galactic-scale
  environments in extreme-mass-ratio binaries}.
\newblock {\em Phys. Rev. D}, 107(8):084027, 2023.

\bibitem{Maselli:2021men}
Andrea Maselli, Nicola Franchini, Leonardo Gualtieri, Thomas~P. Sotiriou,
  Susanna Barsanti, and Paolo Pani.
\newblock {Detecting fundamental fields with LISA observations of gravitational
  waves from extreme mass-ratio inspirals}.
\newblock {\em Nature Astron.}, 6(4):464--470, 2022.

\bibitem{Cornish:2008zd}
Neil~J. Cornish.
\newblock {Detection Strategies for Extreme Mass Ratio Inspirals}.
\newblock {\em Class. Quant. Grav.}, 28:094016, 2011.

\bibitem{Babak:2009ua}
Stanislav Babak, Jonathan~R. Gair, and Edward~K. Porter.
\newblock {An Algorithm for detection of extreme mass ratio inspirals in LISA
  data}.
\newblock {\em Class. Quant. Grav.}, 26:135004, 2009.

\bibitem{Gair:2004iv}
Jonathan~R. Gair, Leor Barack, Teviet Creighton, Curt Cutler, Shane~L. Larson,
  E.~Sterl Phinney, and Michele Vallisneri.
\newblock {Event rate estimates for LISA extreme mass ratio capture sources}.
\newblock {\em Class. Quant. Grav.}, 21:S1595--S1606, 2004.

\bibitem{Wang:2012xh}
Yan Wang, Yu~Shang, Stanislav Babak, Yu~Shang, and Stanislav Babak.
\newblock {EMRI data analysis with a phenomenological waveform}.
\newblock {\em Phys. Rev. D}, 86:104050, 2012.

\bibitem{Wang:2015kja}
Yan Wang, Gerhard Heinzel, and Karsten Danzmann.
\newblock {First stage of LISA data processing II: Alternative filtering
  dynamic models for LISA}.
\newblock {\em Phys. Rev. D}, 92(4):044037, 2015.

\bibitem{Lynch:2021ogr}
Philip Lynch, Maarten van~de Meent, and Niels Warburton.
\newblock {Eccentric self-forced inspirals into a rotating black hole}.
\newblock {\em Class. Quant. Grav.}, 39(14):145004, 2022.

\bibitem{Isoyama:2021jjd}
Soichiro Isoyama, Ryuichi Fujita, Alvin J.~K. Chua, Hiroyuki Nakano, Adam
  Pound, and Norichika Sago.
\newblock {Adiabatic Waveforms from Extreme-Mass-Ratio Inspirals: An Analytical
  Approach}.
\newblock {\em Phys. Rev. Lett.}, 128(23):231101, 2022.

\bibitem{George:2017pmj}
Daniel George and E.~A. Huerta.
\newblock {Deep Learning for Real-time Gravitational Wave Detection and
  Parameter Estimation: Results with Advanced LIGO Data}.
\newblock {\em Phys. Lett. B}, 778:64--70, 2018.

\bibitem{George:2016hay}
Daniel George and E.~A. Huerta.
\newblock {Deep Neural Networks to Enable Real-time Multimessenger
  Astrophysics}.
\newblock {\em Phys. Rev. D}, 97(4):044039, 2018.

\bibitem{Wei:2019zlc}
Wei Wei and E.~A. Huerta.
\newblock {Gravitational Wave Denoising of Binary Black Hole Mergers with Deep
  Learning}.
\newblock {\em Phys. Lett. B}, 800:135081, 2020.

\bibitem{Wang:2019zaj}
He~Wang, Shichao Wu, Zhoujian Cao, Xiaolin Liu, and Jian-Yang Zhu.
\newblock {Gravitational-wave signal recognition of LIGO data by deep
  learning}.
\newblock {\em Phys. Rev. D}, 101(10):104003, 2020.

\bibitem{Jin:2023ahl}
Shang-Jie Jin, Yu-Xin Wang, Tian-Yang Sun, Jing-Fei Zhang, and Xin Zhang.
\newblock {Rapid identification of time-frequency domain gravitational wave
  signals from binary black holes using deep learning}.
\newblock 5 2023.

\bibitem{Ruan:2021fxq}
Wen-Hong Ruan, He~Wang, Chang Liu, and Zong-Kuan Guo.
\newblock {Rapid search for massive black hole binary coalescences using deep
  learning}.
\newblock {\em Phys. Lett. B}, 841:137904, 2023.

\bibitem{Zhang:2022xuq}
Xue-Ting Zhang, Chris Messenger, Natalia Korsakova, Man~Leong Chan, Yi-Ming Hu,
  and Jing-dong Zhang.
\newblock {Detecting gravitational waves from extreme mass ratio inspirals
  using convolutional neural networks}.
\newblock {\em Phys. Rev. D}, 105(12):123027, 2022.

\bibitem{Zhao:2022qob}
Tianyu Zhao, Ruoxi Lyu, He~Wang, Zhoujian Cao, and Zhixiang Ren.
\newblock {Space-based gravitational wave signal detection and extraction with
  deep neural network}.
\newblock {\em Commun. Phys.}, 6(1):212, 2023.

\bibitem{Zhao:2023ncy}
Tianyu Zhao, Yue Zhou, Ruijun Shi, Zhoujian Cao, and Zhixiang Ren.
\newblock {DECODE: DilatEd COnvolutional neural network for Detecting
  Extreme-mass-ratio inspirals}.
\newblock 8 2023.

\bibitem{Gair:2007bz}
Jonathan~R. Gair, Ilya Mandel, and Linqing Wen.
\newblock {Time-frequency analysis of extreme-mass-ratio inspiral signals in
  mock LISA data}.
\newblock {\em J. Phys. Conf. Ser.}, 122:012037, 2008.

\bibitem{Gair:2008ec}
Jonathan~R. Gair, Ilya Mandel, and Linqing Wen.
\newblock {Improved time-frequency analysis of extreme-mass-ratio inspiral
  signals in mock LISA data}.
\newblock {\em Class. Quant. Grav.}, 25:184031, 2008.

\bibitem{babak2020lisa}
S~Babak and A~Petiteau.
\newblock Lisa data challenge manual.
\newblock Technical report, Tech. Rep. LISA-LCST-SGS-MAN-002, APC Paris, 2020.

\bibitem{Tinto:2003vj}
Massimo Tinto, Frank~B. Estabrook, and J.~W. Armstrong.
\newblock {Time delay interferometry with moving spacecraft arrays}.
\newblock {\em Phys. Rev. D}, 69:082001, 2004.

\bibitem{Vallisneri:2004bn}
Michele Vallisneri.
\newblock {Synthetic LISA: Simulating time delay interferometry in a model
  LISA}.
\newblock {\em Phys. Rev. D}, 71:022001, 2005.

\bibitem{Katz:2022yqe}
Michael~L. Katz, Jean-Baptiste Bayle, Alvin J.~K. Chua, and Michele Vallisneri.
\newblock {Assessing the data-analysis impact of LISA orbit approximations
  using a GPU-accelerated response model}.
\newblock {\em Phys. Rev. D}, 106(10):103001, 2022.

\bibitem{Barack:2003fp}
Leor Barack and Curt Cutler.
\newblock {LISA capture sources: Approximate waveforms, signal-to-noise ratios,
  and parameter estimation accuracy}.
\newblock {\em Phys. Rev. D}, 69:082005, 2004.

\bibitem{Chua:2017ujo}
Alvin J.~K. Chua, Christopher~J. Moore, and Jonathan~R. Gair.
\newblock {Augmented kludge waveforms for detecting extreme-mass-ratio
  inspirals}.
\newblock {\em Phys. Rev. D}, 96(4):044005, 2017.

\bibitem{Babak:2006uv}
Stanislav Babak, Hua Fang, Jonathan~R. Gair, Kostas Glampedakis, and Scott~A.
  Hughes.
\newblock {'Kludge' gravitational waveforms for a test-body orbiting a Kerr
  black hole}.
\newblock {\em Phys. Rev. D}, 75:024005, 2007.
\newblock [Erratum: Phys.Rev.D 77, 04990 (2008)].

\bibitem{Moore:2014lga}
C.~J. Moore, R.~H. Cole, and C.~P.~L. Berry.
\newblock {Gravitational-wave sensitivity curves}.
\newblock {\em Class. Quant. Grav.}, 32(1):015014, 2015.

\end{thebibliography}

\end{document}